\documentclass[letter,scriptaddress,twocolumn, prl,showkeys]{revtex4}

\usepackage{bm}

	\usepackage{amsmath}
	\usepackage{makeidx}
	\usepackage{amsfonts}
	\usepackage[ansinew]{inputenc}
	\usepackage[usenames,dvipsnames]{pstricks}
	\usepackage{subfigure}
	\usepackage{epsfig}
	\usepackage{pst-grad} 
	\usepackage{pst-plot} 
	\usepackage[colorlinks,hyperindex]{hyperref}
	\usepackage{tikz}
	\hypersetup
	{
		colorlinks,%
		citecolor=black,%
		linkcolor=black,%
		urlcolor=black,%
	}

\usepackage{gensymb}

\newcommand{\beq}{\begin{equation}}
\newcommand{\eeq}{\end{equation}}

\makeindex

\begin{document}

\title{Giga-Gauss scale quasistatic magnetic field generation in an 'escargot' target}

\author{Ph.Korneev}
\email{korneev@theor.mephi.ru}
\address{NRNU MEPhI, Moscow 115409, Russian Federation and University of Bordeaux, CNRS, CEA, CELIA, 33405 Talence, France}
\author{E. d'Humi\`eres, V. Tikhonchuk}
\address{University of Bordeaux, CNRS, CEA, CELIA, 33405 Talence, France}



\begin{abstract}
{A simple setup for the generation of ultra-intense quasistatic magnetic fields, based on the generation of electron currents with a predefined geometry in a curved 'escargot' target, is proposed and analysed. Particle-In-Cell simulations and qualitative estimates show that giga-Gauss scale magnetic fields may be achieved with existent laser facilities. The described mechanism of the strong magnetic field generation may be useful in a wide range of applications, from laboratory astrophysics to magnetized ICF schemes. 
}
\end{abstract}

\keywords{Magnetic field, laboratory astrophysics, $\theta-$pinch, laser-plasma interaction, inverse currents.}

\maketitle





\section{Introduction}

Generation of strong magnetic fields in laboratory conditions attracts much interest as it may be used in various of applications, such as astrophysical studies, Inertial Confinement Fusion (ICF) schemes, magnetic field interaction with atoms and particles, etc. For the laboratory production of magnetic fields of the order of hundreds of kilo-Gauss, pulsed magnetic sources can be used. Nowadays, modern laser facilities provide new possibilities for the one-shot generation of intense magnetic fields up to tens of Mega-Gauss (see, i.e. \cite{ryutov-ass11, fujioka-sr12,yoneda-prl12}), in a volume of $\sim1$ mm$^3$ and in a time scale of several ns.


In the present letter, we propose a novel scheme for the production of intense magnetic fields, based on the generation of intense currents  in a target of a special 'escargot' geometry \cite{korneev-arxiv14}. The sceme can be understood from Fig.\ref{history}(A1), a laser pulse propagates inside the target from the left through the 'window'. As a result of grazing incidence and the geometry of the target, various physical phenomena are involved into the interaction process. We mention here laser pulse reflection from plasma surface, electron surface guiding effect \cite{nakamura-prl04}, and return current generation. A combination of these three effects in the considered target geometry results in a strong magnetic field generation. 
The curvature of the inner target surface provides (i) the continuous laser propagation along the surface and a high total absorption, (ii) acceleration of electrons along the curved target surface, (iii) the return current with a target-defined curvature.
The electron guiding along the target inner surface (ii), known as the electron surface acceleration mechanism, is described in \cite{nakamura-prl04}. The effect was experimentally confirmed \cite{li-prl06}, for laser intensities $\sim10^{18}$W/cm$^2$, and studied numerically in different geometries \cite{psikal-pop10, nakamura-pop07}. It was found, that the accelerated electrons produce strong currents along the surface, with the corresponding magnetic fields. Another important effect, not mentioned in \cite{korneev-arxiv14}, is the return current generation (iii) in a target material. It results in generation of higher quasistatic magnetic field of the opposite direction. The field amplitude may reach a giga-Gauss level, with the characteristic life time of the order of at least several ps. This quasistationary magnetic field may deflect the surface electron guiding, and, moreover, produce higly magnetized electron off-surface flows.  In this letter, with Particle-in-Cell simulations we show an example of giga-Gauss scale magnetic field generation and analyse the physical origin of this effect.

The letter is organized as follows: first, we present the results of the Particle-In-Cell simulations, then discuss the magnetic field structure and its origin, and finally conclude.

\section{Particle-in-Cell simulations for an example of the magnetic field generation in an 'escargot' target}
\begin{figure}
\begin{center}
\hspace*{-1cm}%
  \begin{tikzpicture}[x=0.54\linewidth,y=0.5\linewidth]
    \path
%
    (1,0) node{\includegraphics[width=0.58\linewidth]{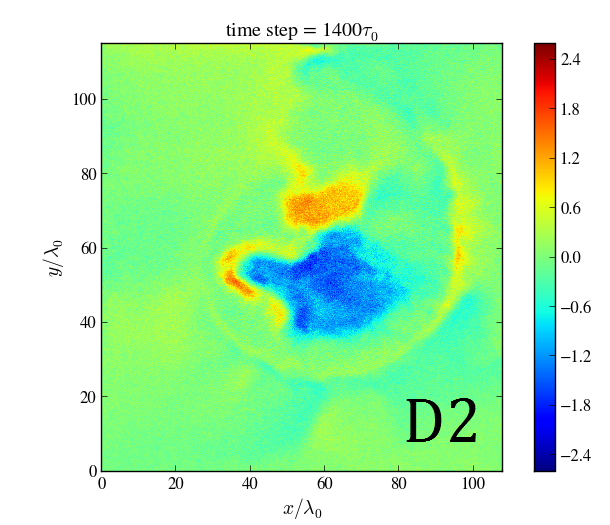}}
    (1,1) node{\includegraphics[width=0.58\linewidth]{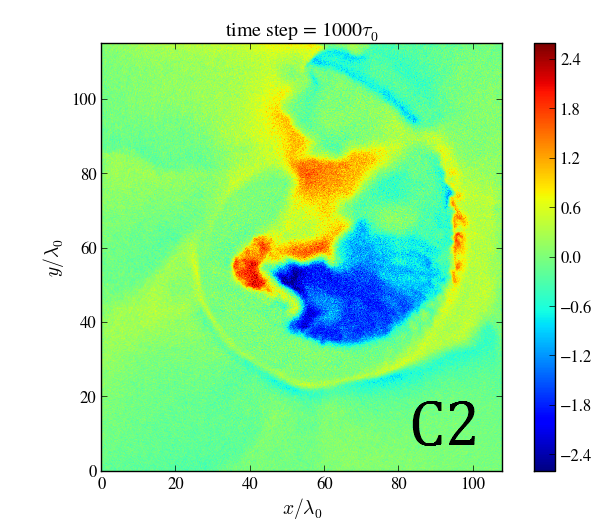}}
    (0.99,2) node{\includegraphics[width=0.61\linewidth]{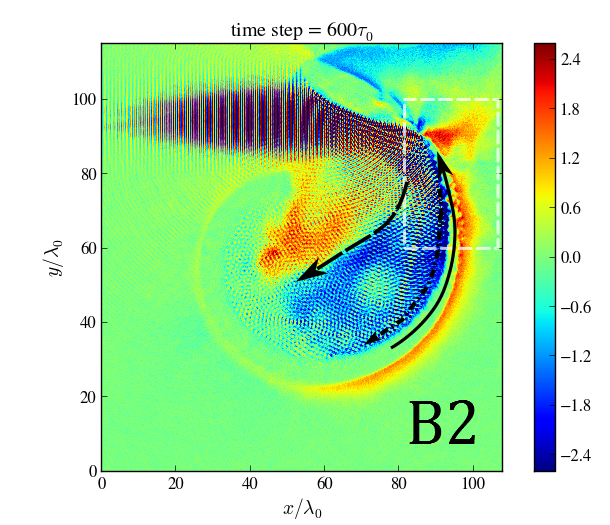}}
    (1,3) node{\includegraphics[width=0.58\linewidth]{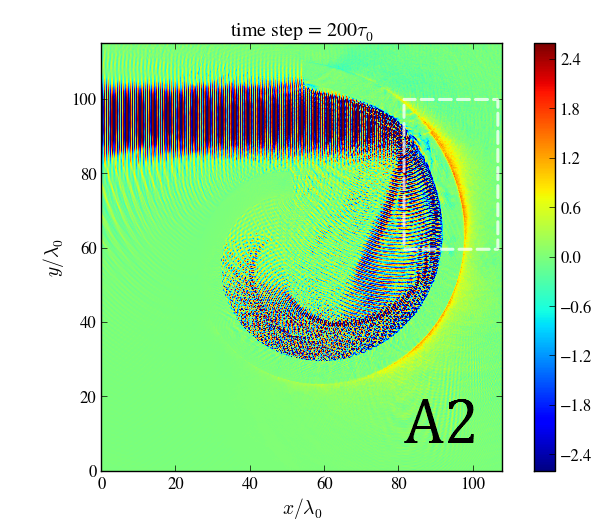}}

    (0,0) node{\includegraphics[width=0.6\linewidth]{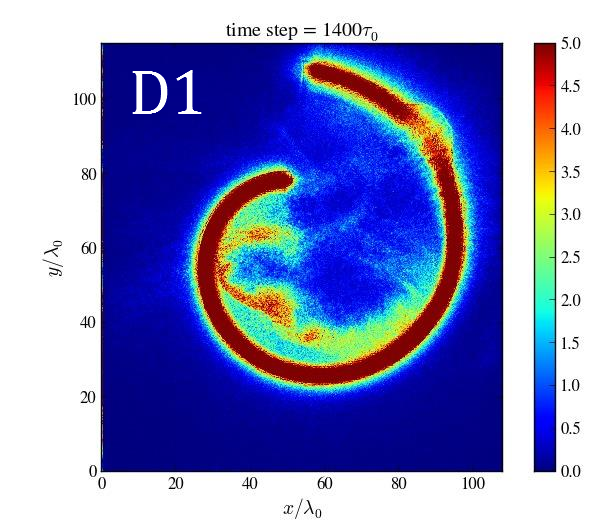}}
    (0,1) node{\includegraphics[width=0.6\linewidth]{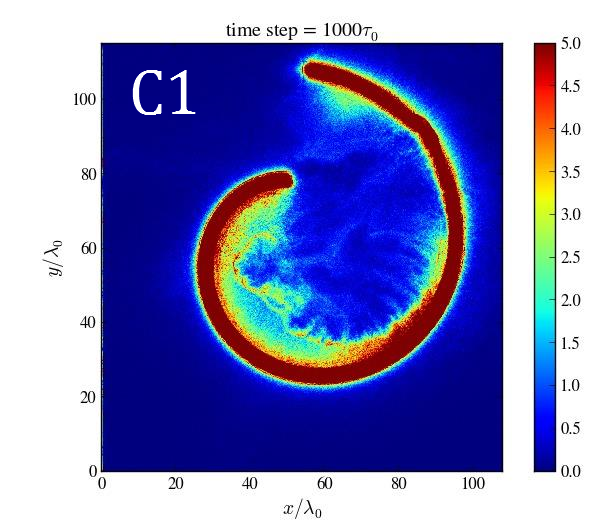}}
    (0,2) node{\includegraphics[width=0.6\linewidth]{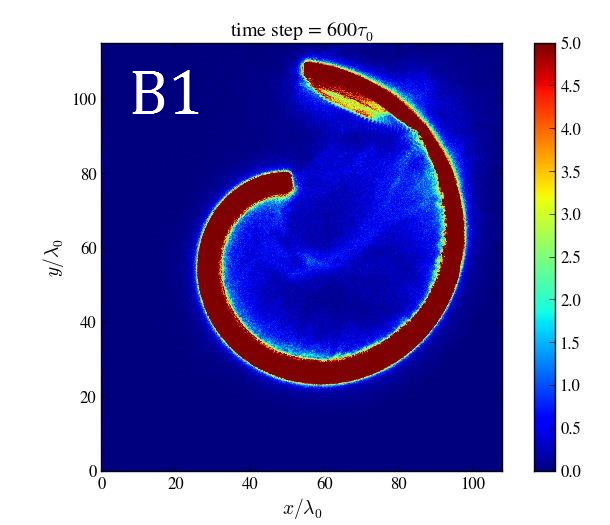}}
    (0,3) node{\includegraphics[width=0.6\linewidth]{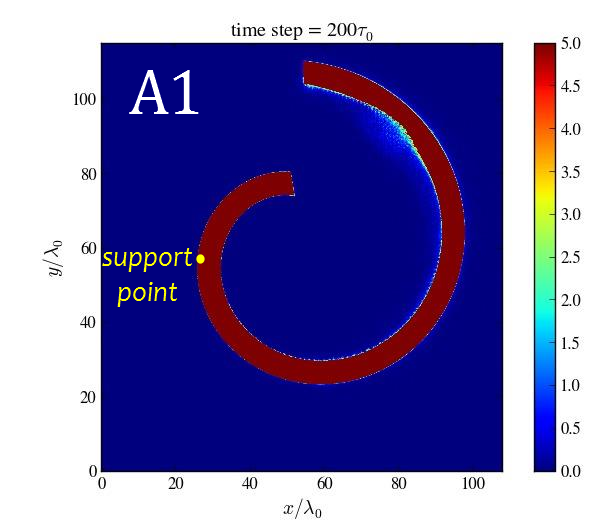}}

    (1.01,1.86) node {$\bm{\langle - \vec j_s \rangle}$}
     (1.04,2.08) node {$\bm{\langle - \vec j_3 \rangle}$}
     (1.28,1.8) node {$\bm{\langle - \vec j_r \rangle}$}

;
  \end{tikzpicture}
\caption{Electron density (panels A1-D1), and magnetic field $B_z$ (panels A2-D2) at different time moments: $0.62,~1.9,~3.1,~4.3$ ps correspondingly for A1,B1,C1,D1 and A2,B2,C2,D2, for the run (a). Electron density is shown in the units of $n_c=1.3\times10^{21}$cm$^{-3}$, and is cut on the value of $6.5\times10^{21}$cm$^{-3}$ magnetic field is shown in the units of $1.16\times10^{8}$Gauss, so that maximum value of $2.6$ in the colorbar corresponds to $3\times10^{8}$Gauss. Axis '$z$' is directed to the viewer. In panel B2 the black dashed arrow along the target inner surface indicates surface guided electrons, the black solid arrow corresponds to the electrons which produce reverse current, and the long dashed arrow shows electron motion, which is deflected by the magnetic field, already formed inside the target.}
\label{history}
\end{center}
\end{figure}

To produce a strong curved solenoid-like current, a special helix, or 'escargot' target geometry is proposed, see Fig.\ref{history}(A1). Our simulations are two-dimensional, so the real 3D shape of the target is somewhat like a disclosed deformed cylinder. This geometry allows us to make use of the plasma mirror effect \cite{Mourou-optics-rel-RevModPhys06pdf.pdf}, with a mirror of a prescribed geometry, and to predefine the currents directions. 
The discontinuity of the target is nesessary for the propagation of the laser pulse inside the hollow. The target shape is analytically defined by 
\beq
r(\theta)=r_0\left(1+\frac{\delta r}{r_0}\frac{\theta}{2\pi}\right),~~~\theta\in\left( 0, 2\pi \right),
\label{target}
\eeq
where $\theta=0$ corresponds to the upper direction of the vertical axis in Fig.\ref{history}, $\delta r$ and $r_0$ are the parameters, which values are defined below. We examine laser-target interaction with 2D3V Particle-In-Cell code PICLS \cite{Sentoku-jcp08}. 



Although several runs with different laser and target parameters were performed to prove the robustness of the proposed setup, here, we describe only one of them. The laser intensity in it was $I_0=5\times10^{19}$W/cm$^{2}$, the laser wavelength was $\lambda_0=0.93~\mu$m, and the pulse duration was $\tau_0=1.6$ps. The target, defined by (\ref{target}), with $r_0=43~\mu$m, $\delta r=28~\mu$m, was composed by two layers of the material with ion charge $Z=79$, ion mass $m_i^{(1)}=197 m_p$ of the inner $1~\mu$m width layer, and $m_i^{(2)}=20m_i^{(1)}$ of the outer $2~\mu$m width layer (sublayer), where $m_p$ is the proton mass. The ion density was $n_i=2\times10^{20}$cm$^{-3}$. Electrons with masses $m_e$ had the density $n_e=Zn_i=12n_c$, where $n_c=1.3\times10^{21}$cm$^{-3}$ is the critical electron density. The more massive outer sublayer was introduced to decrease the target explosion time. The matter was presented as fully ionized cold ions, 1 particle per a cell, and 100keV hot electrons, 79 particles per a cell. The simulation box was $2160\times2304$ cells, or approximately $100\times110~\mu$m. The resolution in time was $0.16$ fs. 

\begin{figure}
\hspace*{-1cm}%
  \begin{tikzpicture}[x=0.31\linewidth,y=0.42\linewidth]
    \path
	(0,0) node{\includegraphics[width=0.4\linewidth]{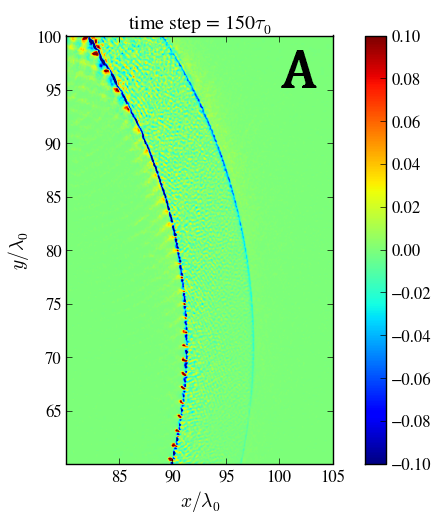}}

%
	(1,0) node{\includegraphics[width=0.4\linewidth]{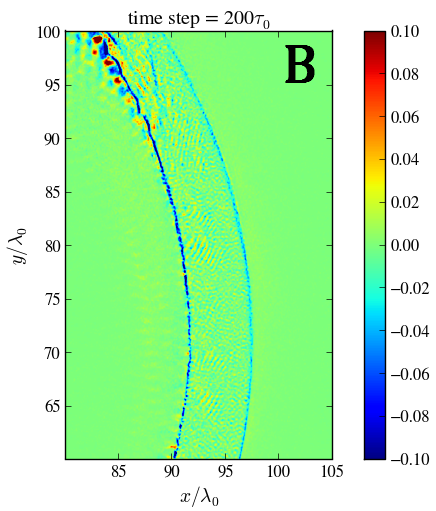}}
	(2,0) node{\includegraphics[width=0.4\linewidth]{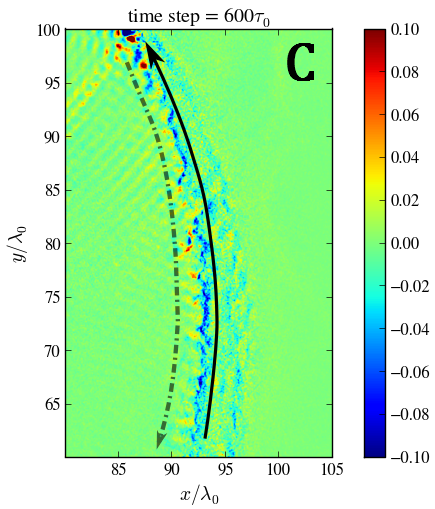}}
     (1.7,-0.3) node {$\bm{\langle - \vec j_s \rangle}$}
     (2.09,0.3) node {$\bm{\langle -\vec j_r \rangle}$}
;
  \end{tikzpicture}
\caption{Current density $j_y$ in the frame region, indicated in Fig.\ref{history}(A2,B2), in subsequent time moments $0.46,~0.62,~1.9$ ps. Negative more intense current is responsible for negative $B_z$ in Fig.\ref{history}(A2,B2,C2,D2). In right panel the dot-dashed arrow shows electrons, directly accelerated by the laser pulse, the solid arrow shows electrons which produce inverse current. }
\label{history2}
\end{figure}

The results of the PIC simulations are shown by several subsequent snapshots of electron density and magnetic field $B_z$ during the interaction process in Fig.\ref{history}. As it is seen from the snapshots for the magnetic field in Fig. \ref{history}(A2-D2), even though a large part of the laser pulse energy is absorbed by the target, a sufficient part of it follows the geometry of the 'escargot' chamber, which works as a plasma mirror of the specified form. To reveal the origin of the magnetic fields, the current structure snapshots are shown in Fig.\ref{history2}. The total current is composed both from the direct electrons along the surface, ponderomotively accelerated by the surface guided mechanism \cite{nakamura-prl04}, and from the reverse current, generated in the target in order to compensate the charge accumulation caused by escaped electrons. The total current produces $B_z$ magnetic field, which effects the further electron dynamics. In the upper part of the target ($\sim 0..10\degree$), where the laser propagation is almost grazing, the surface guiding effect is more important. Below, in the main region of the laser pulse reflection, incident angle increases up to $\sim40..45\degree$. Under these conditions the surface guiding mechanism still works  \cite{nakamura-prl04}, and the corresponding direct current is seen in Fig.\ref{history2}(A) as a thin layer of positive current along the surface. The inverse current (negative in Fig.\ref{history2}), appears to be more stable and posesses a higher value in comparison to the direct one. While laser propagates further, it heats more and more electrons on the inner surface of the target. These electrons escape from the surface, creating the local positive charge. This is the nature of a \textit{return current feeding} by the laser propagation along curved target surface. The direction and the amplitude of the magnetic field inside the cavity is defined by the total current distribution, but as the escaped electrons are spread and the reverse current is bound inside the target material, the later plays the decisive role for the magnetic field generation. Note that the resulting magnetic structure posess a $\theta-$pinch-like geometry, which is known to be stable, and thus its life time is defined generally by the target life time.
Later on, see Fig.\ref{history}(C1,C2), Target Normal Sheath Acceleration (TNSA) mechanism \cite{mora-prl03} of ion acceleration comes into play, pulling ions from the target inner surface into the inner region. This secondary effect leads to an additional compression of the magnetic field by the plasma pressure.


\begin{figure}
\includegraphics[width=9cm]{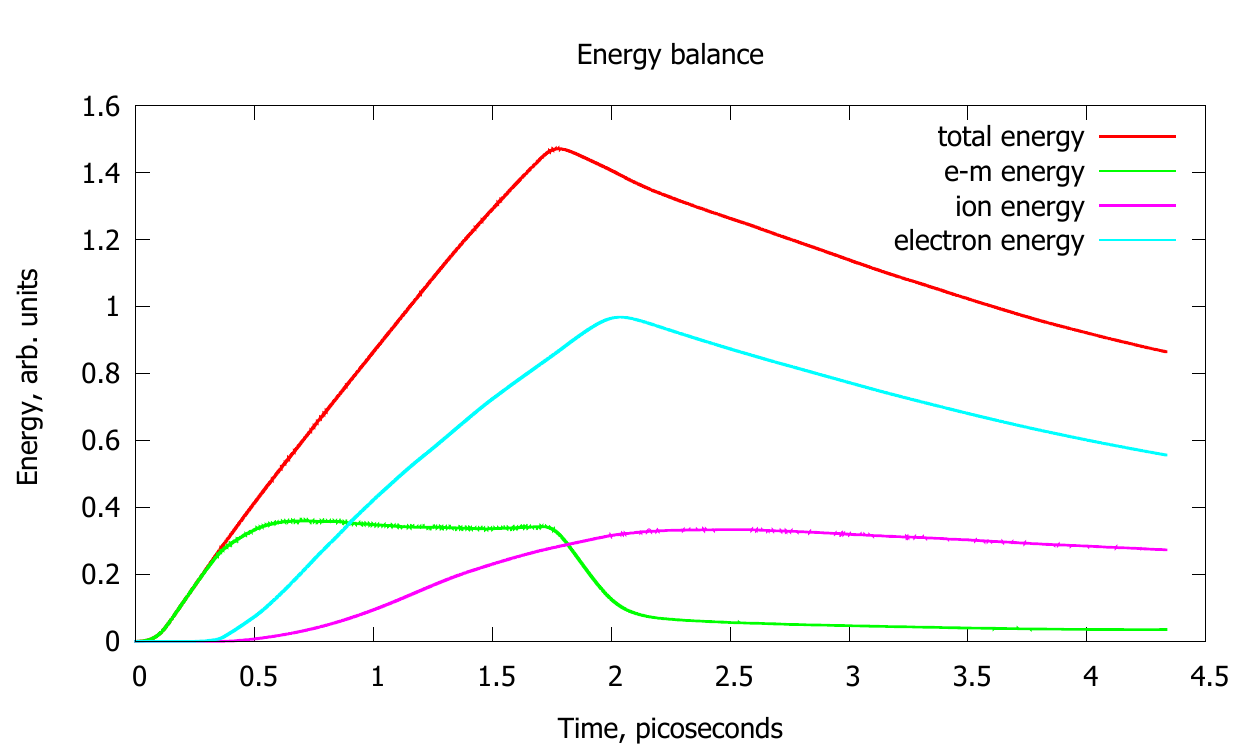}
\caption{Energy balance during the interaction. After 2 ps, when the laser pulse is gone, the electromagnetic energy is composed only by the magnetic field energy, which has the order of $5-7\%$ of the maximum (total laser pulse) energy.}
\label{energy}
\end{figure}

The energy conversion efficiency from laser to the remaining magnetic field may be estimated from the energy balance in the PIC simulations, shown in Fig.\ref{energy}. In this plot, the total energy growth is approximately linear during the time, when laser pulse is entering, and after it ends, energy gradually decreases, because the most energetic particles leave the simulation box. Initially, the electromagnetic energy also grows linearly, but approximately at 0.4 ps, when the laser reaches the target, its growth stops due to the electron heating. In the contrary, during the laser interaction with the target, the electron energy grows, reaching the value of the order of $\approx 60\%$ of the total energy at the end of the laser pulse (about 1.5 ps). Also a sufficient part of the total energy is transmitted to ions due to the TNSA effect. Electromagnetic energy remains of the same order of $\approx35$\% of the total laser energy during the laser-target interaction process, but when the laser pulse ends it does not decrease to the zero value. There is a small remaining part, left after 2 ps, which corresponds to the energy of the residual magnetic field. From Fig.\ref{energy} it is seen, that this magnetic energy contains about $5-7\%$ of the total laser pulse energy.

\section{Discussion}

The interaction and the magnetic field development consist on several subsequent processes. These are electron heating by the laser pulse, currents generation, magnetic structure formation, and TNSA effect. We discuss first the current structure and estimate the corresponding magnetic fields. 
Although as we mention, the main role in the magnetic field generation plays the reverse current, the direct electron current along the surface, which in our case may be generated in the large surface area may also be important. For estimate of the the magnetic field amplitude, the Amp\`ere's law may be used,
\beq
\nabla\times \vec B = \frac{4\pi}{c}\vec j +\frac{1}{c}\frac{\partial \vec E}{\partial t},
\label{ampere}
\eeq
where $\vec B$ and $\vec E$ are the magnetic and the electric fields, $\vec j$ is current, and $c$ is the light velocity. For the estimate of the quasi-static magnetic field we average (\ref{ampere}) over time $\tau_{av}$, which is much longer than both the laser period $\tau_{av}\gg\omega^{-1}$ and the electron plasma period $\tau_{av}\gg\omega_e^{-1}$. After this, the time derivative item $\langle\partial\vec E/\partial t\rangle$ in (\ref{ampere}) becomes very small and can be omitted. This means, that the average currents $\langle\vec j\rangle$ form a self-consistent structure with the magnetic fields $\langle\vec B\rangle$, and that the fast creation of the bounding electric field during electron heating on the inner target surface does not play an important role in the quasi-static $\langle\vec B\rangle$ generation. According to the current structure, shown in Fig.\ref{history2}(C), it is necessary to estimate both the direct surface current $\langle\vec j_s\rangle$ and the reverse one $\langle\vec j_r\rangle$. It is convenient first to estimate the total number of the escaping electrons from the surface and their average energy. 

For relativistic laser intensities, the characteristic electron energy $T_e$ is defined by the ponderomotive scaling \cite{wilks-prl92}, which for $I_0=5\times10^{19}$W/cm$^2$ gives $T_e\sim 4$ MeV. The number of escaped electrons  in a stationary regime $N_{esc}$ can be found from the self-consistent solution of Poisson-Boltzmann problem \cite{POYE-metalic_charge_target_v6}, which gives for the focal radii $r_{f}\sim10~\mu$m, $N_{esc}\sim 6\times10^{12}$ electrons. This number gives the total electric charge and the return current $\langle\vec j_r\rangle$, in the the target during the action of the laser pulse $\tau_0$. According to the numerical simulations, the number of electrons along the surface, which defines the current $\langle\vec j_s\rangle$ contains about $\kappa\sim0.1..0.2\lesssim1$ of the total electron number $N_{esc}$ for moderate oblique angles \cite{nakamura-prl04}. It can not considerably decrease the magnetic field, created by the return current $\langle\vec j_r\rangle$.
According to (\ref{ampere}), the generated by the total current $\langle \vec j \rangle=\langle\vec j_s\rangle+\langle\vec j_r\rangle$ magnetic field is $\langle B_z \rangle\approx0.5\times10^9$ Gauss, which is in a reasonable agreement with the values, obtained in the PIC simulations, see Fig.\ref{history}.

When the magnetic field is formed, the electrons forming the surface current $\langle\vec j_s\rangle$, as well as all other escaping electrons, may be  deflected. The Larmor radius for electrons with an energy of $4$ MeV in a magnetic field of the value from Fig.\ref{history}(B2) $2\times10^8$ Gauss, is much smaller than the target size: $r_L\lesssim 1~\mu$m. This means, that electrons can not penetrate into the magnetized region. However, they can propagate along the edge of the strong magnetic field region, and form the third current $\langle \vec j_3 \rangle$, indicated in Fig.\ref{history}(B2) with the long-dashed arrow. In the given geometry, this electrons propagate approximately through the middle of the empty space inside the target, separating it and forming a dipole-like magnetic field distribution. Variation of the $\delta r/r_0$ parameter in (\ref{target}) and the laser propagation direction may affect the shape of the magnetic field structure: for a large $\delta r/r_0$, the current $-\langle \vec j_3 \rangle$ is ejected outside the cavity and the generated magnetic field may take a monopole shape. 


The longitudinal magnetic structure lives long after the end of the laser pulse. Moreover, we observe an interesting effect of a compression of the magnetic field by the plasma expanding inside the cavity. This process is relatively slow, and may be qualitatively described by the pressure balance between the magnetic field and the hot plasma, 
\begin{gather}
 \frac{B_z^2}{8\pi} \approx n_e T_e,
\label{equilibrium}
\end{gather}
which is usual for $\theta-$pinch configurations. Taking from Fig.\ref{history} the electron density $n_e\sim 10^{21}$cm$^{-3}$ and temperature $T_e\sim 4$MeV, we get the magnetic field value $B_z\sim 4\times10^{8}$ Gauss. This value is twice greater, than the value in Fig.\ref{history}(D2), probably because of the electron cooling after the laser pulse ends.

At these times, inside the target cavity we observe an interaction of the low density hot magnetized plasma (in the blue region with the  negative $\langle B_z\rangle$ in Fig.\ref{history}(C2,D2)) and more dense cool magnetized plasma (in the yellow region with the positive $\langle B_z\rangle$  in Fig.\ref{history}(C2,D2)). The low density plasma region is formed along the more heated right part of the target. There the dense hot collisionless surface plasma cannot be rapidly magnetized and remains for some time separated from the magnetized region. In the left part of the target, plasma is relatively cold, but the magnetic field, generated by $\langle \vec j_3 \rangle$ is strong. There a considerable part of the surface plasma may be magnetized, see Fig.\ref{history}(C1,C2). Finally, the ablated plasma comes to the equilibrium (\ref{equilibrium}) with the magnetic field inside the cavity. This quasi-static stage is defined mainly by the target explosion time, so that the more dense materials may be preferable to increase the time of life of the generated magnetic structure.

\section{Conclusions and perspectives}


We presented two-dimentional calculations for the high quasistationary magnetic field generation mechanism with intense laser pulses. For a possible experimental realization, the question of the role of 3D effects arises. The $\theta-$pinch type magnetic structure, observing in our simulations, is expected to be stable at the time scale of several ps. The required laser intensity of $5\times10^{19}$ W/cm$^2$ and the pulse duration of $1.6$ ps could be achieved with the laser energy of the order of $1$ kJ. This is the scale of the installations as PETAL, ORION, OMEGA EP, and FLEX. 

It is known \cite{POYE-metalic_charge_target_v6}, that reverse currents depend strongly on the experimental realization, including a holder position on the target. For the considered setup, we marked a possible 'support point', which should be placed so, that not to strongly change current distribution during the interaction.


%
%

\begin{figure}
  \begin{tikzpicture}[x=0.32\linewidth]
    \path
    (0,0) node{\includegraphics[scale=0.22]{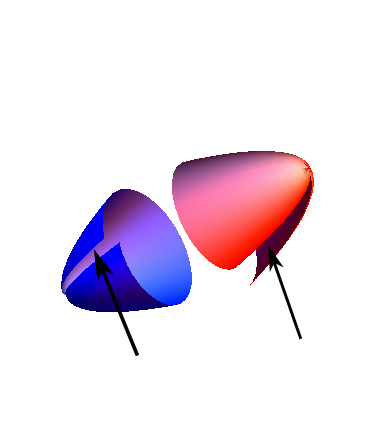}}
    (1,0) node{\includegraphics[scale=0.22]{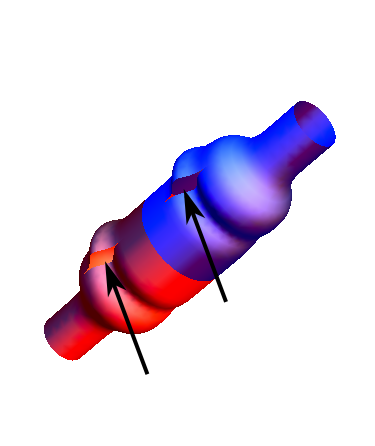}}
    (2,0) node{\includegraphics[scale=0.22]{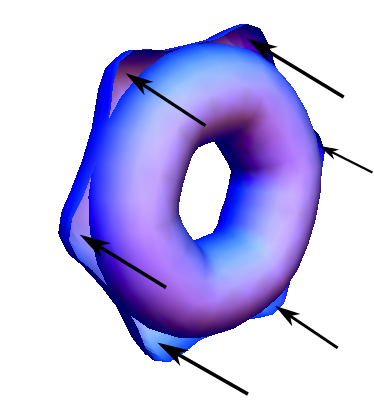}}

    (0,-1.5) node { a }
    (1,-1.5) node { b }
    (2,-1.5) node { c }

;
  \end{tikzpicture}
\caption{Examples of target geometries for the experimental applications of the considered effect: a -- two cone-like 'escargot' targets for collisions of magnetized plasmas; b -- magnetic trap geometry; c -- microtocamak geometry. Black arrows show laser pulses directions.}
\label{examples}
\end{figure}

The robustness of the considered mechanism of magnetic fields generation, was confirmed in several runs with different target and laser parameters. A strong return current in a predefined geometry of the target can be a flexible and effective experimental tool.
The proposed scheme for the magnetic field generation may be used in a variety of applications, such as laboratory astrophysics experiments, neutron production, different aspects of ICF, i.e. electron magnetic collimation \cite{perez-prl11}, etc. Several variations are presented in Fig\ref{examples}. 
In laboratory astrophysics applications, a magnetized plasma may be used for the studies of collisionless shocks and magnetic reconnection phenomena \cite{yamada-rmp10}. Depending on laser and target parameters it may be possible to create magnetized plasmas, propagating in the opposite directions, with different orientations of the magnetic field. This can avoid the magnetic compression \cite{korneev-pop14}, which is considered as not typical in space plasmas. A possible setup for the reconnection studies is shown in Fig.\ref{examples}a. Inside a single 'escargot' target (\ref{target}), with adjusted laser and target parameters, reconnection phenomena may also take place, as it follows from the structure of the magnetic field in Fig.\ref{history}.
An interesting application may be micrometer-scale magnetic traps, shown in Fig.\ref{examples}b and \ref{examples}c. Target sizes and magnetic field values may be of the order of interest for neutron production or magnetized fusion schemes \cite{Hasegawa-prl86}. For the magnetic field $\langle B \rangle \approx100$MGauss, and the trap radius is about $50~\mu$m, it can contain protons with energies $\sim30$ MeV, and $\alpha-$particles with energies of the order of $\sim10$ MeV. 

%


In conclusion, in the present Letter a novel scheme of laser-assisted production of intense magnetic fields is proposed. It is based on the generation of the intense currents, which propagating along the curved surface. As a first "proof-of-principle" example, with the 'escargot'-like target we demonstrated a simple quasi-static $\theta-$pinch type magnetic structure. More complex magnetic field micro-structures may be produced with more complicated target geometries.


\section*{ACKNOWLEDGMENTS}

Authors greatly appreciate usefull discussions with S.Fujioka, A.Poy\'e and J.J.Santos. The work is in part supported by the French National
funding agency ANR within the project SILAMPA, and it was granted access to the HPC resources of CINES under the allocation 2014-056129 made by GENCI (Grand Equipement National de Calcul Intensif).


%




\end{document}